\documentclass[
superscriptaddress,
nobibnotes,
notitlepage,
amsmath,amssymb,
aps,
twocolumn,
prl,
floatfix,
]{revtex4-1}

\usepackage[T1]{fontenc}
\usepackage{aas_macros}
\usepackage{graphicx}
\usepackage{dcolumn}
\usepackage{bm, color, colortbl}
\usepackage{lipsum, babel}

\usepackage{comment}
\usepackage{lineno}

\usepackage[citecolor=blue]{hyperref}

\usepackage{orcidlink}

\definecolor{linkcolor}{rgb}{0.6,0,0}
\definecolor{citecolor}{rgb}{0,0,0.75}
\definecolor{urlcolor}{rgb}{0.12,0.46,0.7}
\hypersetup{colorlinks,linkcolor=linkcolor,citecolor=citecolor,urlcolor=urlcolor}  

\usepackage{ctable}
\usepackage{array}
\usepackage{graphicx}
\usepackage[large]{subfigure}
\usepackage{amssymb, amsmath}
\usepackage[amssymb]{SIunits}
\usepackage[normalem]{ulem}
\usepackage{aas_macros}
\usepackage{natbib}
\renewcommand\section[1]{\emph{#1}.---}

\usepackage{xcolor}
\usepackage{booktabs}
\usepackage{multirow} 

\newcolumntype{C}[1]{>{\centering\arraybackslash\hspace{0pt}}p{#1}}
\newcolumntype{B}{>{\columncolor{blue!30}[.5\tabcolsep]}C{2.6cm}}
\begin{document}

\preprint{APS/123-QED}

\newcommand{\Hunit}{{\rm km}\ {\rm s}^{-1}\rm{Mpc}^{-1}}
\newcommand{\Frank}[1]{\textcolor{blue}{(FJQ: #1)}}
\newcommand{\hun}{\,\mathrm{km}\,\mathrm{s}^{-1}\mathrm{Mpc}^{-1}}
\newcommand{\KW}[1]{\textcolor{purple}{(KW: #1)}}
\newcommand{\bds}[1]{\textcolor{cyan}{(BDS: #1)}}
\newcommand{\AC}[1]{\textcolor{orange}{(AC: #1)}}
\newcommand{\ns}[1]{\textcolor{magenta}{(NS: #1)}}
\newcommand{\FG}[1]{\textcolor{teal}{(FG: #1)}}
\newcommand{\kw}[1]{\textcolor{purple}{#1}}
\newcommand{\gsf}[1]{\textcolor{green}{(GSF: #1)}}
\newcommand{\iac}[1]{\textcolor{pink}{(IAC: #1)}}
\newcommand{\sptshort}{{SPT-3G M2PM}} 


\title{Unified and consistent structure growth measurements \\from joint ACT, SPT and \textit{Planck} CMB lensing}

\author{Frank J. Qu \,\orcidlink{0000-0001-7805-1068}}
\email{jq247@cantab.ac.uk}
\affiliation{Kavli Institute for Particle Astrophysics and Cosmology, Stanford University, 452 Lomita Mall, Stanford, CA, 94305, USA}
\affiliation{Department of Physics, Stanford University, 382 Via Pueblo Mall, Stanford, CA, 94305, USA}
\affiliation{Kavli Institute for Cosmology Cambridge, Madingley Road, Cambridge CB3 0HA, UK}

\author{Fei Ge\,\orcidlink{0000-0002-3833-8133}}
\affiliation{Kavli Institute for Particle Astrophysics and Cosmology, Stanford University, 452 Lomita Mall, Stanford, CA, 94305, USA}
\affiliation{Department of Physics, Stanford University, 382 Via Pueblo Mall, Stanford, CA, 94305, USA}
\affiliation{Department of Physics \& Astronomy, University of California, One Shields Avenue, Davis, CA 95616, USA}
\affiliation{Department of Physics, California Institute of Technology, Pasadena, CA, 91125, USA}

\author{W.~L.~Kimmy Wu\,\orcidlink{0000-0001-5411-6920}}
\affiliation{Kavli Institute for Particle Astrophysics and Cosmology, Stanford University, 452 Lomita Mall, Stanford, CA, 94305, USA}
\affiliation{SLAC National Accelerator Laboratory, 2575 Sand Hill Road, Menlo Park, California 94025, USA}
\affiliation{Department of Physics, California Institute of Technology, Pasadena, CA, 91125, USA}
\author{Irene Abril-Cabezas\,\orcidlink{0000-0003-3230-4589}}
\affiliation{DAMTP, Centre for Mathematical Sciences, University of Cambridge, Wilberforce Road, Cambridge CB3 0WA, UK}
\affiliation{Kavli Institute for Cosmology Cambridge, Madingley Road, Cambridge CB3 0HA, UK}
\author{Mathew~S.~Madhavacheril}
\affiliation{Department of Physics and Astronomy, University of Pennsylvania, 209 South 33rd Street, Philadelphia, PA 19104, USA}
\author{Marius~Millea\,\orcidlink{0000-0001-7317-0551}}
\affiliation{Department of Physics \& Astronomy, University of California, One Shields Avenue, Davis, CA 95616, USA}
\author{Zeeshan~Ahmed}
\affiliation{Kavli Institute for Particle Astrophysics and Cosmology, Stanford University, 452 Lomita Mall, Stanford, CA, 94305, USA}
\affiliation{SLAC National Accelerator Laboratory, 2575 Sand Hill Road, Menlo Park, California 94025, USA}
\author{Ethan~Anderes}
\affiliation{Department of Statistics, University of California, One Shields Avenue, Davis, CA 95616, USA}
\author{Adam~J.~Anderson\,\orcidlink{0000-0002-4435-4623}}
\affiliation{Fermi National Accelerator Laboratory, MS209, P.O. Box 500, Batavia, IL, 60510, USA}
\affiliation{Kavli Institute for Cosmological Physics, University of Chicago, 5640 South Ellis Avenue, Chicago, IL, 60637, USA}
\affiliation{Department of Astronomy and Astrophysics, University of Chicago, 5640 South Ellis Avenue, Chicago, IL, 60637, USA}
\author{Behzad~Ansarinejad}
\affiliation{School of Physics, University of Melbourne, Parkville, VIC 3010, Australia}
% ...
\collaboration{ACT + SPT-3G Collaborations}
\noaffiliation


\date{\today}

\begin{abstract}

We present the tightest cosmic microwave background (CMB) lensing constraints to date on the growth of structure by combining CMB lensing measurements from the Atacama Cosmology Telescope (ACT), the South Pole Telescope (SPT) and \textit{Planck}. Each of these surveys individually provides lensing measurements with similarly high statistical power, achieving signal-to-noise ratios of approximately 40. 
The combined lensing bandpowers represent the most precise CMB lensing power spectrum measurement to date with a signal-to-noise ratio of 61 and an amplitude of $A_\mathrm{lens}^\mathrm{recon} = 1.025 \pm 0.017$ with respect to the theory prediction from the best-fit CMB \textit{Planck}-ACT cosmology. The bandpowers from all three lensing datasets, analyzed jointly, yield a $1.6\%$ measurement of the parameter combination $S_8^\mathrm{CMBL} \equiv \sigma_8\,(\Omega_m/0.3)^{0.25} = 0.825^{+0.015}_{-0.013}$. Including Dark Energy Spectroscopic Instrument (DESI) Baryon Acoustic Oscillation (BAO) data improves the constraint on the amplitude of matter fluctuations to $\sigma_8 = 0.829 \pm 0.009$ (a $1.1\%$ determination). When combining with uncalibrated supernovae from \texttt{Pantheon+}, we present a $4\%$ sound-horizon-independent estimate of $H_0=66.4\pm2.5\,\mathrm{km\,s^{-1}\,Mpc^{-1}} $. The joint lensing constraints on structure growth and present-day Hubble rate are fully consistent with a $\Lambda$CDM model fit to the primary CMB data from \textit{Planck} and ACT. While the precise upper limit is sensitive to the choice of data and underlying model assumptions, when varying the neutrino mass sum within the $\Lambda\mathrm{CDM}$ cosmological model, the combination of primary CMB, BAO and CMB lensing drives the probable upper limit for the mass sum towards lower values, comparable to the minimum mass prior required by neutrino oscillation experiments.

\end{abstract}

\maketitle
\section{\label{sec:level1}Introduction}
Lensing of the cosmic microwave background (CMB), 
the deflection of CMB photon paths by intervening large-scale structure, has emerged as a highly robust probe of the mass distribution.

    Building on lensing measurements from the satellite-based missions \textit{Wilkinson Microwave Anisotropy Probe } (\textit{WMAP}) \citep{Smith_2007} and \textit{Planck} \citep{p2014,p2016,p2020}, and from ground-based surveys such as ACT \citep{Das_2011,Sherwin_2017} and SPT \citep{van_Engelen_2012,Story_2015,Wu:2019hek,SPT:2019fqo,SPT:2023jql}, the measurement of CMB lensing has been advanced to the regime of precision cosmology.
    Notably, direct structure growth measurements via CMB lensing are consistent with the predictions of the $\Lambda$CDM model conditioned on the primary CMB measurements \footnote{For example, this can be seen from \textit{Planck} \citep{Planck:2018_cosmo_params,Rosenberg:2022sdy} or from a combination of \textit{Planck} large scales with ACT arcminute-scale measurements \citep{act_0,act_1,act_2}}.

 In this work, we combine the latest results from {\it Planck}~PR4~\cite{Planck:2020olo}, ACT DR6~\cite{ACT:2023dou,ACT:2023kun,ACT:2023ubw} and SPT-3G MUSE analysis on the main-field-2-year polarization-only data~\citep[][hereafter \sptshort]{SPT-3G:2024atg}.
 These three analyses use data from successively narrower fields and with lower noise levels, yet they achieve comparable lensing power spectrum signal-to-noise ratios. 
 The consistency between these independent measurements, despite their distinct observational strategies, makes their agreement, and consequently the joint constraints presented here, compelling.

 We infer the amplitude of structure growth, specifically via a parameter combination of linear matter power fluctuation ($\sigma_8$) and the fractional matter energy density ($\Omega_m$), $S^{\mathrm{CMBL}}_8\equiv \sigma_8(\Omega_m/0.3)^{0.25}$, which CMB lensing is most sensitive to in the redshift range $z\approx0.9$--$5$ and the physical wavenumber range $k \approx 0.05 \sim 0.3 \ \mathrm{Mpc^{-1}}$. This is complementary to the constraints from galaxy surveys \cite{DES:2021bvc, DES:2021vln, DES:2021wwk, LSSTDarkEnergyScience:2022amt, KiDS:2020suj, Heymans:2020gsg, Li:2023tui, Dalal:2023olq, Wright:2025xka}, which are sensitive to lower redshifts and larger $k$ values (smaller physical scales).

We then include baryon acoustic oscillation (BAO) observations, which act as a probe of $\Omega_m$, thereby allowing us to constrain $\sigma_8$ separately.  We also determine the Hubble constant, $H_0$, using two approaches: one relying on the sound horizon scale to which BAO is sensitive, and one based on the matter-radiation equality scale to which CMB lensing is sensitive. Finally, we use the combination of the ACT$+$\textit{Planck} primary CMB, BAO from DESI and our lensing measurements to revisit cosmological limits on the neutrino mass sum.

\section{Data}
We briefly describe the datasets and the external likelihoods used in this work.

\begin{figure}[t!]
    \centering
    \includegraphics[width=\columnwidth]{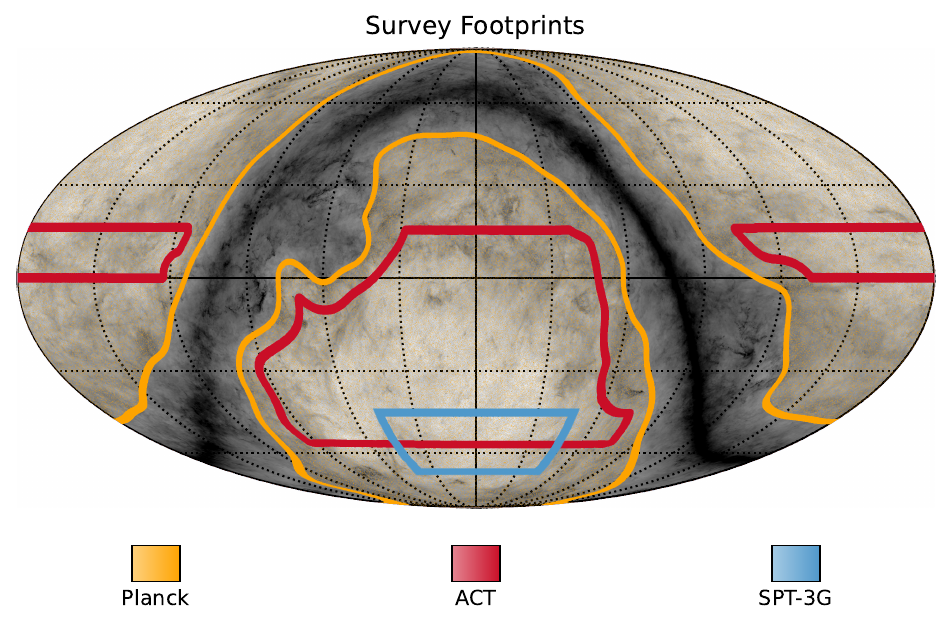}
    \caption{Mollweide projection showing the sky coverage of ACT DR6 (red), \textit{Planck} (orange) and \sptshort\ (blue). ACT DR6 covers $23\%$ of the sky, \sptshort\ covers $3.5\%$ and they overlap across $2.1\%$ of the sky. \textit{Planck} PR4 covers $67\%$ of the sky. The grayscale background is a Galactic dust map from {\it Planck} \cite{1502.01588}. }
    \label{fig:footprint}
\end{figure}

\textbf{CMB lensing spectra:} We employ the CMB lensing spectrum measurements from ACT DR6 \cite{ACT:2023dou,ACT:2023kun,ACT:2023ubw}, \textit{Planck} PR4~\cite{Carron_2022} and \sptshort~\citep{SPT-3G:2024atg} with their respective survey footprints shown in Fig.~\ref{fig:footprint}.

The ACT DR6 lensing reconstruction~\cite{ACT:2023dou,ACT:2023kun,ACT:2023ubw} (red in Fig.~\ref{fig:footprint}) covers $23\%$ of the sky and is signal-dominated on lensing scales with multipoles $L<150$. The lensing spectrum is measured with a signal-to-noise ratio of $43$ using a cross-correlation-based quadratic estimator that is insensitive to the modeling of instrumental noise~\citep{Madhavacheril_2021,Atkins_2023}. In this analysis, we use the extended ACT DR6 multipole range of $40\leq{L}\leq1300$.

The \textit{Planck} PR4 lensing analysis \cite{Planck:2020olo} reconstructs lensing with the quadratic estimator using the reprocessed PR4 \texttt{NPIPE} CMB maps. It covers $67\%$ of the sky (orange in Fig. \ref{fig:footprint}) and is signal-dominated below $L \approx 70$. The lensing spectrum is measured with a signal-to-noise ratio of $42$.

The \sptshort\ lensing measurement~\citep{SPT-3G:2024atg} covers $3.5\%$ of the sky (blue in Fig.~\ref{fig:footprint}). This analysis derives lensing information from CMB polarization maps using data collected with the SPT-3G camera during the 2019 and 2020 observing seasons. The analysis employs the Marginal Unbiased Score Expansion (MUSE) method \citep{PhysRevD.105.103531,millea2022improvedmarginalunbiasedscore} to infer CMB lensing and unlensed EE power spectra jointly. 
The lensing spectrum is signal-dominated for lensing multipoles $L < 240$ and is measured with a signal-to-noise ratio of~$38$.

\textbf{BAO:} We use the results from DESI data release 2 (DR2), consisting of BAO measured from more than 14 million galaxies and quasars, as well as the DESI  Lyman-$\alpha$ BAO \citep{desicollaboration2025desidr2resultsi,desicollaboration2025desidr2resultsii} as our baseline baryonic acoustic oscillation combination. In the \hyperref[app.variation]{\textit{supplementary material}} ~\cite{SupplementalMaterial}, we show that using alternative BAO datasets 
yields consistent results with those obtained using DESI DR2 BAO.

\textbf{Supernovae:} For constraints on the Hubble constant that do not rely on the sound horizon, we include `uncalibrated' Type Ia supernovae from \texttt{Pantheon+}~\citep{2022ApJ...938..110B} as our baseline sample, although we also compare with \texttt{UNION3} \citep{rubin2024unionunitycosmology2000} and \texttt{DESY5} \citep{descollaboration2024darkenergysurveycosmology}. Here, `uncalibrated' indicates that only the relation between the apparent magnitudes of Type Ia supernovae and their redshifts is employed—without anchoring their absolute magnitudes—and hence using these data  cannot yield a determination of $H_0$.

\textbf{Primary CMB:} When constraining the neutrino mass sum, we add the primary CMB power spectrum measurement from \textit{Planck} PR3 (including \texttt{high-$\ell$ TTTEEE}, \texttt{lowl T} and \texttt{SRoll2 EE} \citep{Planck:2019nip, Delouis:2019bub}) and ACT DR6 \citep{act_0, act_1, act_2} (hereafter P-ACT), following the procedure in \citep{act_1, act_2} to combine the two datasets. 
The comparisons of our measurements to the primary CMB are also done with respect to the P-ACT best-fit cosmology of \citep{act_2} that we henceforth denote as \textsf{CMB}.

\section{Likelihood}\label{seclikelihood}
We build a Gaussian likelihood\footnote{The likelihood is publicly available at \url{https://github.com/qujia7/spt_act_likelihood}.} from the CMB lensing bandpowers of ACT DR6 \cite{ACT:2023dou,ACT:2023kun,ACT:2023ubw}, \textit{Planck} PR4 \cite{Planck:2020olo} and \sptshort\ \citep{SPT-3G:2024atg}:
\begin{equation}\label{eq:likelihood}
    -2 \ln \mathcal{L} \propto \sum_{bb'}\begin{bmatrix}\Delta \hat{C}_b^{\kappa_A\kappa_A} \\ \Delta \hat{C}_b^{\kappa_P\kappa_P} \\ \Delta \hat{C}_b^{\kappa_S\kappa_S} \end{bmatrix} \mathbb{C}^{-1}_{b b'} \begin{bmatrix}\Delta \hat{C}_{b'}^{\kappa_A\kappa_A} \\ \Delta \hat{C}_{b'}^{\kappa_P\kappa_P}\\ \Delta \hat{C}_{b'}^{\kappa_S\kappa_S}\end{bmatrix},
\end{equation}
where $\Delta \hat{C}_b^{\kappa_i\kappa_i}$ ($i\in[A,P,S]$) are the residuals between observed and theory CMB lensing spectra for ACT DR6 (A), \textit{Planck} PR4 (P), and \sptshort~(S). The covariance matrix $\mathbb{C}_{b b'}$ includes auto-covariances from simulations and cross-covariances between experiments (see \hyperref[app.likelihood_details]{supplementary material}). Cross-correlations between ACT-SPT ($\lesssim15\%$) and \textit{Planck}-SPT ($\lesssim10\%$) are small due to limited sky overlap and the different weighting of temperature versus polarization in the reconstructions.

We infer cosmological parameters using MCMC with \texttt{Cobaya}~\citep{2021JCAP...05..057T}, evaluating fiducial lensing bandpowers with the \texttt{class\_sz} emulator~\citep{Bolliet:2017lha,Bolliet:2023eob} for $\Lambda$CDM (with $\sum m_\nu = 60$ meV) and \texttt{CAMB}~\cite{Lewis:1999bs, Howlett:2012mh} for $\Lambda$CDM$+\Sigma m_\nu$ models. Priors follow ACT DR6~\citep{ACT:2023kun} (Table~\hyperref[table:priors]{I} of the \hyperref[app.prior]{\textit{supplementary material}}).

\section{Results}\label{sec:results}

\begin{figure*}[t]
    \centering
    \includegraphics[width=\linewidth]{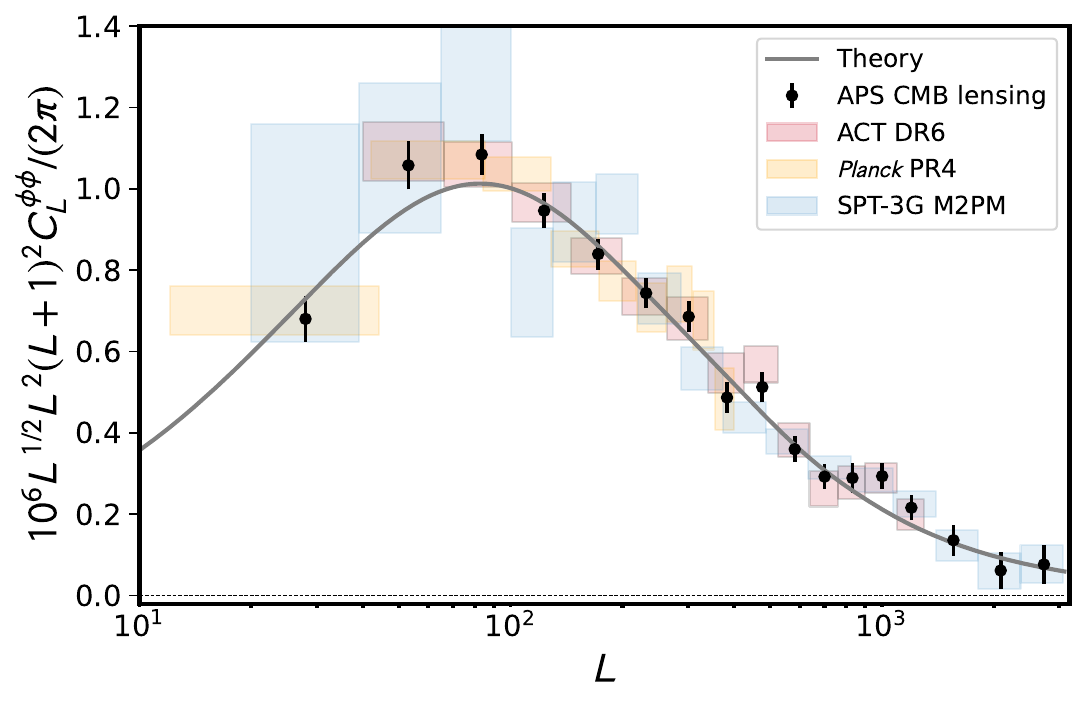}
    \caption{We present the combined lensing bandpowers from the three surveys in black. In the background we  show the \textit{Planck} lensing bandpowers from PR4 \texttt{NPIPE} analysis in orange,  the ACT DR6 lensing potential power spectrum bandpowers in red, and the lensing bandpowers from {\sptshort}  in blue. The gray line shows the theory prediction from the best-fit cosmology of the \textsf{CMB} likelihood. Note that we have applied an additional $L^{1/2}$ scaling over that usually used to display bandpowers to enhance visually the small scales. }
    \label{fig:combclkk}
\end{figure*}

\subsection{\textbf{SPT-3G lensing-only constraints on structure growth}}

We present CMB lensing only constraints using \sptshort. In \cite{SPT-3G:2024atg}, the amplitudes of CMB lensing and structure growth are derived simultaneously with CMB lensing and unlensed CMB EE bandpowers. In this work, since we aim to assess consistency across the three CMB lensing datasets, in the following we first report the lensing-only constraints from \sptshort.

We estimate the lensing amplitude parameter $A_\mathrm{lens}^\mathrm{recon}$ from \sptshort\ by fitting the SPT lensing bandpower measurements to a theory lensing power spectrum based on the best-fit $\Lambda$CDM model from $\textsf{CMB}$, allowing the amplitude of this lensing power spectrum to be a free parameter in our fit. 

We find ${A_\mathrm{lens}^\mathrm{recon}}=1.033\pm0.026\quad(68\%\, \mathrm{C.L.})$, in good agreement with the $\textsf{CMB}$ $\Lambda$CDM  prediction (i.e., $A_\mathrm{lens}^\mathrm{recon}=1$), with a PTE $\chi^2$ of $17\%$ \footnote{we note that these results are consistent with the one combining \textit{Planck} primary CMB and \sptshort\ lensing in~\citep{SPT-3G:2024atg}.}.

Analyzing only the SPT lensing bandpowers, we obtain a $1.9\%$ constraint on structure growth given by 
\begin{equation}
    S^{\mathrm{CMBL}}_8=0.827\pm0.016~~~(68\% \ \mathrm{C.L.},  \ \textrm{\sptshort}).
\end{equation}

The high-precision small-scale lensing bandpowers from \sptshort\ are highly complementary to those from ACT and {\it Planck}, which obtain higher precision on larger scales. The combination of all three datasets enables the tightest constraints on $S_8^{\mathrm{CMBL}}$ to date. With good agreement on $S_8^{\mathrm{CMBL}}$ between \sptshort, ACT DR6 and {\it Planck} PR4 lensing (see Table \ref{tab:params1}), we proceed in the next section to obtain results from the likelihood-level combination of the three CMB lensing measurements.

\begin{table}[ht]
\begin{tabular}{l  c  c  c    }
\hline\hline
Experiment & $S^{\mathrm{lens}}_8$ & $\sigma_8$ & $\Omega_m$ \\ \hline
A & $0.830 \pm 0.020$ & - & - \\
P & $0.809\pm0.022$ & - &-\\
S & $0.827 \pm 0.016$ & - & - \\
APS & $0.825^{+0.015}_{-0.013}$ & - & - \\
A+BAO & $0.826 \pm 0.015$ & $0.827 \pm 0.014$ & $0.298 \pm 0.008$ \\
P+BAO & $0.808\pm0.018$  & $0.811\pm0.016$ &  $0.295\pm0.008$\\
S+BAO  & $0.830 \pm 0.012$ & $0.831 \pm 0.012$ & $0.298 \pm 0.008$ \\
APS+BAO & $0.829\pm{0.009}$ & $0.829 \pm 0.009$ & $0.300\pm0.007$\\
\hline
\end{tabular}
\caption{Cosmological parameter measurements from the various lensing experiment combinations. We use A, P and S as shorthands for CMB lensing with ACT DR6, \textit{Planck} PR4 and \sptshort, respectively.}
\label{tab:params1}
\end{table}

\subsection{ACT+SPT+\textit{Planck} (APS) joint constraints on structure growth}

In Fig.~\ref{fig:combclkk}, we show the individual lensing spectra from ACT DR6 (red), \textit{Planck} PR4 (orange) and \sptshort~(blue). The joint lensing bandpowers, which {are}
signal-dominated at $L\lesssim240$, {are} obtained by performing an amplitude fit on the bins between the three surveys against a theoretical lensing power spectrum predicted from the $\textsf{CMB}$ best-fit $\Lambda$CDM model, in a similar way to a Bayesian linear regression \footnote{The joint bandpowers are only used to calculate the lensing amplitude, the SNR and for the visual representation in Fig.~\ref{fig:combclkk}. We use Eqn.~(\ref{eq:likelihood}) for all the parameter constraints in the rest of this paper.}.

The joint bandpowers have a lensing amplitude of 
\begin{equation}    A_\mathrm{lens}^\mathrm{recon}=1.025\pm0.017\ (68\% \ \mathrm{C.L.},  \ \ {\rm APS}),
\end{equation}
with a signal-to-noise ratio of 61, making this the most precise CMB lensing power spectrum measurement to date and in excellent agreement with the primary CMB predictions within the $\Lambda$CDM model (we obtain similar $A_\mathrm{lens}^\mathrm{recon}=1.010\pm0.016$ when comparing to the \textit{Planck} best fit cosmology).

\begin{figure}[t]
    \centering
    \includegraphics[width=\columnwidth]{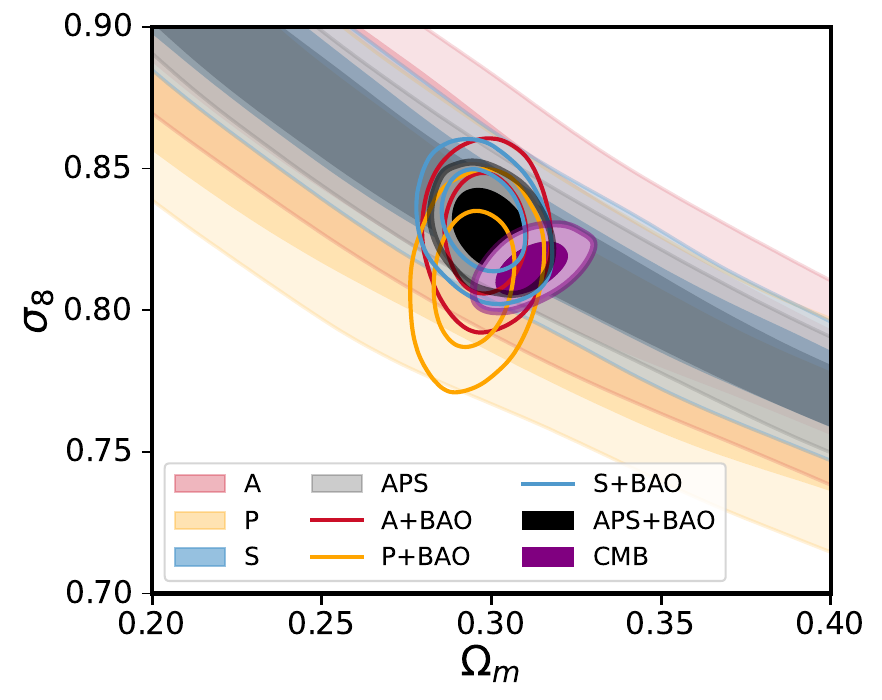}
    \caption{Marginalized posteriors in the $\sigma_8$--$\Omega_m$ plane for ACT DR6 (red), \textit{Planck} PR4 (orange), \sptshort\ (blue) and APS (black) CMB lensing measurements. Filled contours in the background show lensing‐only results, except for the black filled contour which represents APS+BAO. Non-filled contours (outlined) show results when including BAO data, which further breaks degeneracies in structure growth. The purple contours show the \textsf{CMB} prediction for a $\Lambda$CDM model. Each data set is shown with their $68\%$ and $95\%$ confidence limits.}
    \label{fig:comb_constraint}
\end{figure}

We measure $S^\mathrm{CMBL}_8$, which is the parameter combination best constrained by CMB lensing within the  $\Lambda$CDM model, to $1.6\%$:
\begin{equation}
    S^{\mathrm{CMBL}}_8= 0.825^{+0.015}_{-0.013}~(68\% \ \mathrm{C.L.},  \ {\rm APS}).
\end{equation}

We can compare this result with the value expected from an extrapolation of the \textsf{CMB} data constraints within a $\Lambda$CDM cosmology,  $S^\mathrm{CMBL}_8=0.823\pm0.010$; this is fully consistent with our direct measurement \footnote{Here the neutrino mass sum is fixed to 60~meV which is close to the minimum mass allowed in the normal hierarchy by constraints from neutrino oscillations. We later explore the impact of marginalizing over the neutrino mass sum.}.

These CMB lensing measurements provide information about a three-dimensional volume comprising the amplitude of matter fluctuations $\sigma_8$, the matter density $\Omega_m$ and the Hubble constant $H_0$. The inclusion of BAO data provides additional background  information on the expansion history that helps break parameter degeneracies. This enables comparisons of $\sigma_8$ inferred from other probes such as cosmic shear and the primary CMB. With the addition of DESI BAO, we find
\begin{equation}
        \sigma_8= 0.829 \pm 0.009 \  (68\% \ \mathrm{C.L.},  \ {\rm APS}+ {\rm BAO}).
\end{equation}
This $1.1\%$ measurement of $\sigma_8$ is consistent within $1.2\sigma$ with the value inferred from \textsf{CMB}, as can be seen in the marginalized constraints in Fig.~\ref{fig:comb_constraint}. We note that this measurement is the most precise determination of $\sigma_8$  from either galaxy or CMB lensing to date. (See also Fig.~\hyperref[fig:sigma8]{4} in the \hyperref[app.variation]{\textit{supplementary material}}.) We also achieve a competitive constraint on $S_8\equiv\sigma_8(\Omega_m/0.3)^{0.5}$, the parameter combination best measured by cosmic shear, obtaining $S_8=0.828\pm0.012$.

\begin{figure}[t]
    \centering
    \includegraphics[width=\columnwidth]{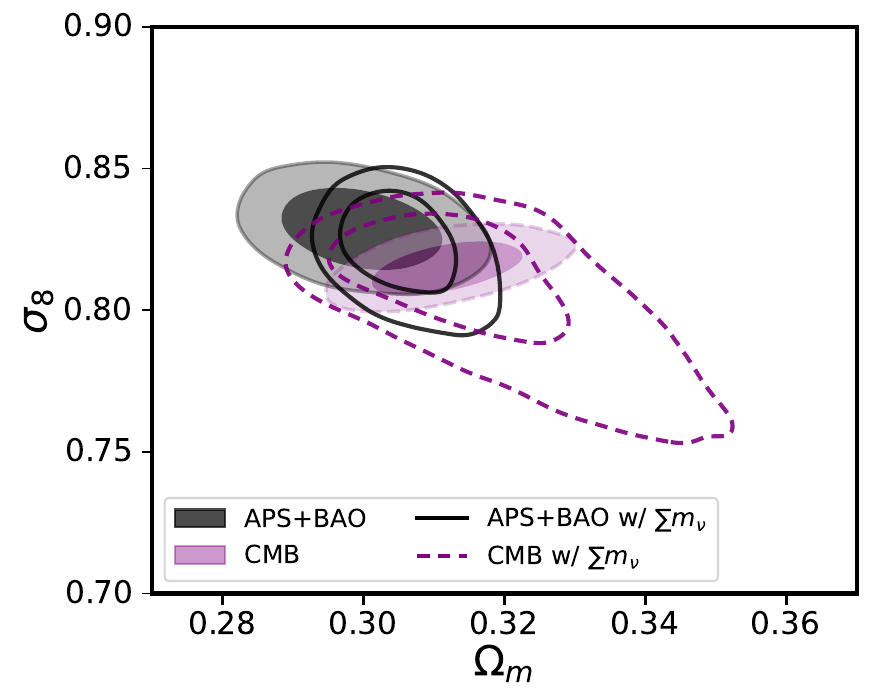}
    \caption{Marginalized posteriors in the $\sigma_8$--$\Omega_m$ plane for APS CMB lensing + BAO (filled black) and the \textsf{CMB} prediction for a $\Lambda$CDM model (purple). Allowing the sum of the neutrino masses to vary results in the open contours. Each data set is shown with their $68\%$ and $95\%$ confidence limits. }
    \label{mnu_free_growth}
\end{figure}

Our lensing measurements are robust to the assumptions of the model---even in the presence of extensions that impact structure growth. The lensing constraints are only slightly weakened when we marginalize over neutrino mass;  in this case we obtain $S^\mathrm{CMBL}_8=0.818^{+0.017}_{-0.013}$, which is comparable to the \textsf{CMB} constraint of $S^\mathrm{CMBL}_8=0.818\pm0.015$ under the same model with free $\sum m_\nu$. This robustness stems from the fact that the lensing measurement originates from relatively low redshifts and hence requires minimal extrapolation to $z=0$ (where $S^\mathrm{CMBL}_8$ is evaluated). The same extrapolation effect and degeneracy breaking when including BAO explains how our $\sigma_8$ constraint with BAO, $\sigma_8=0.822\pm0.012$, becomes more competitive than the \textsf{CMB} extrapolation of $\sigma_8= 0.808^{+0.029}_{-0.040}$  as shown in Fig.~\ref{mnu_free_growth}.

\subsection{ACT+SPT+\textit{Planck} joint constraints on the Hubble constant}

We  use our joint CMB lensing measurements to provide an independent constraint on the Hubble constant, \( H_0 \). While BAO observations, combined with a prior on $\Omega_bh^2$ \footnote{BBN is required to calibrate the BAO and break the degeneracy between $r_d$ and $H_0$.}, are sensitive to the expansion history, they exhibit an extended degeneracy between \( H_0 \) and \( \Omega_m \). In contrast, CMB lensing constrains a different degeneracy direction, making it  complementary to BAO. Combining the baryon-drag-scale~($r_d$-)calibrated BAO with CMB lensing, we break parameter degeneracies and obtain  tighter constraints on \( H_0 \) than from BAO-only measurements.

\begin{figure}[t]
    \centering
    \includegraphics[width=\columnwidth]{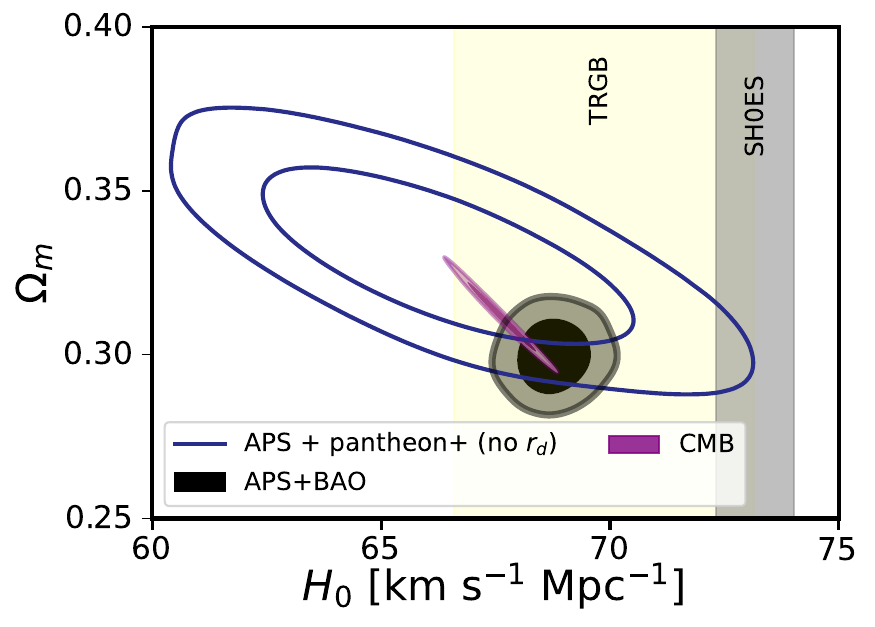}
    \caption{Hubble constant measurements for the combination of CMB lensing and BAO are in filled black contours. The blue open contours show constraints on $H_0$ inferred from the matter--radiation equality scale as opposed to the sound-horizon scale. The $H_0$ measurements with CMB lensing  are consistent with the low expansion rate inferred from the \textsf{CMB} in purple. We also show the  $68\%$ bands of the Cepheid-calibrated direct inference in gray and the TRGB-calibrated direct inference in yellow.}
    \label{fig:comb_expansion_h0} 
\end{figure}

From the combination of the joint CMB lensing, galaxy BAO and the $\Omega_bh^2$ prior in Table~\hyperref[table:priors]{I} of the \hyperref[app.prior]{\textit{supplementary material}}, we obtain a $0.8\%$ constraint on $H_0$:
\begin{align}
        H_0&=68.77\pm 0.53 \hun\\ \nonumber &\hspace{0.05\textwidth}(68\% \ \mathrm{C.L.},  \ \ {\rm APS}+ {\rm\, BAO}).
\end{align}
This result, shown in Fig.~\ref{fig:comb_expansion_h0}, is consistent with the results from the \textsf{CMB} ($H_0=67.62\pm0.50 \hun$) and in around $4\sigma$ tension  with the SH0ES-inferred value of $73.17\pm0.86 \,\, \Hunit$ \cite{Breuval:2024lsv}.

Most of the constraints on $H_0$ using BAO come exclusively from the knowledge of the sound horizon scale $r_d$. Following the method suggested by \citep{Baxter_2020}, we proceed to place sound-horizon-independent constraints on the Hubble constant; these constraints arise instead from the matter-radiation equality scale imprinted in the matter power spectrum to which CMB lensing is sensitive \footnote{Lensing is sensitive to the broadband shape of the matter power spectrum, with the location of the peak determining the scale of the matter-radiation equality.}

Combining our data with uncalibrated supernovae to break the degeneracy between $H_0$ and $\Omega_m$, we find

\begin{align}
        H_0&=66.4^{+2.5}_{-2.8} \hun\\ \nonumber & \hspace{0.05\textwidth}(68\% \ \mathrm{C.L.},  \ \ {\rm APS}+ {\texttt{Pantheon+}}).
\end{align}

We also compute sound-horizon-free measurements using other supernova samples: $H_0 = 64.0^{+2.9}_{-3.5}\hun$ with \texttt{UNION3} and  $ H_0 = 64.2\pm 2.4 \hun$ with \texttt{DESY5}; we note discussions in \citep{efstathiou2025evolvingdarkenergysupernovae} and \citep{vincenzi2025comparingdessn5yrpantheonsn} regarding the \texttt{DESY5} sample. Our sound-horizon-free measurements are consistent with the value of $H_0$ derived from the BAO+APS and primary CMB data. They are also in agreement with the direct distance ladder measurements calibrated using the tip of the red giant branch (TRGB) reported in  \citep{freedman2024statusreportchicagocarnegiehubble} but differ from the SH0ES measurement by $2.5\sigma$   \cite{Breuval:2024lsv}.

\subsection{Neutrino mass}

Massive neutrinos affect structure growth in the Universe after the neutrinos become non-relativistic (e.g., \citep{Lesgourgues_2012}), leading to suppression of the matter power spectrum at the percent level. Since CMB lensing probes the distribution of mass in projection, it is a sensitive probe of the neutrino mass sum. Our baseline constraint uses \textsf{CMB}, BAO and APS CMB lensing, resulting in
\begin{equation}
    \Sigma m_\nu < 0.062 \ \mathrm{eV \ (95\% \ \mathrm{C.L.},  \ \textsf{CMB} + APS + BAO} ) .  \label{eq:mnu}
\end{equation}

The upper limit\footnote{We model neutrinos as a combination of three degenerate equal-mass particles following \citep{Lesgourgues_2012,1612.00021}  consistent with recent cosmological analyses \citep{ACT:2023kun,desicollaboration2025desidr2resultsi,desicollaboration2025desidr2resultsii}. We note that adopting an alternative prescription of one massive and two massless neutrinos, as used by \citep{SPT-3G:2024atg}, yields $\Sigma m_\nu < 0.053 \ \mathrm{eV \ (95\% \ \mathrm{C.L.})}$ using the same data combination as above.} is relatively stable to the primary CMB used (as also noted in \citep{act_2}). Switching to \textit{Planck} NPIPE \texttt{CamSpec}, the upper limit becomes $\sum m_\nu < 0.061\ \mathrm{eV}\ (95\% \ $C.L.). 
While our results show a  preference for a lower $\sum m_\nu$  compared to neutrino oscillation experiments, nominally disfavoring the inverted hierarchy at 
3.3$\sigma$, alternative data combinations or modeling approaches can relax this upper limit, as we will discuss below. 

Our neutrino mass constraints present only modest improvements over others in the literature,  $\sum m_\nu<0.082\ \mathrm{eV}\ (95\% \ $ C.L.) with \textsf{CMB}+AP lensing and DESI DR1 BAO \citep{act_2} and  $\sum m_\nu<0.064\ \mathrm{eV}\ (95\% \ $ C.L.) in the case of \textit{Planck} \textsf{CMB}+AP lensing and DESI DR2 BAO \citep{desicollaboration2025desidr2resultsi,desicollaboration2025desidr2resultsii}.
 References~\cite{Loverde:2024nfi, Lynch:2025ine} argue that such tight constraints arise partly from differences in the inferred matter density, with the matter density $\Omega_mh^2$ inferred from \textsf{CMB}+BAO close to (or even lower) than the mass densities of baryons and cold dark matter inferred from  \textsf{CMB} data \cite{Loverde:2024nfi, Lynch:2025ine}, leaving little room for the neutrino mass density. Reference~\cite{Green:2024xbb} invokes a high lensing amplitude as a key factor in providing unexpectedly tight constraints on the neutrino mass sum.

\begin{figure}[t]
    \centering
    \includegraphics[width=\columnwidth]{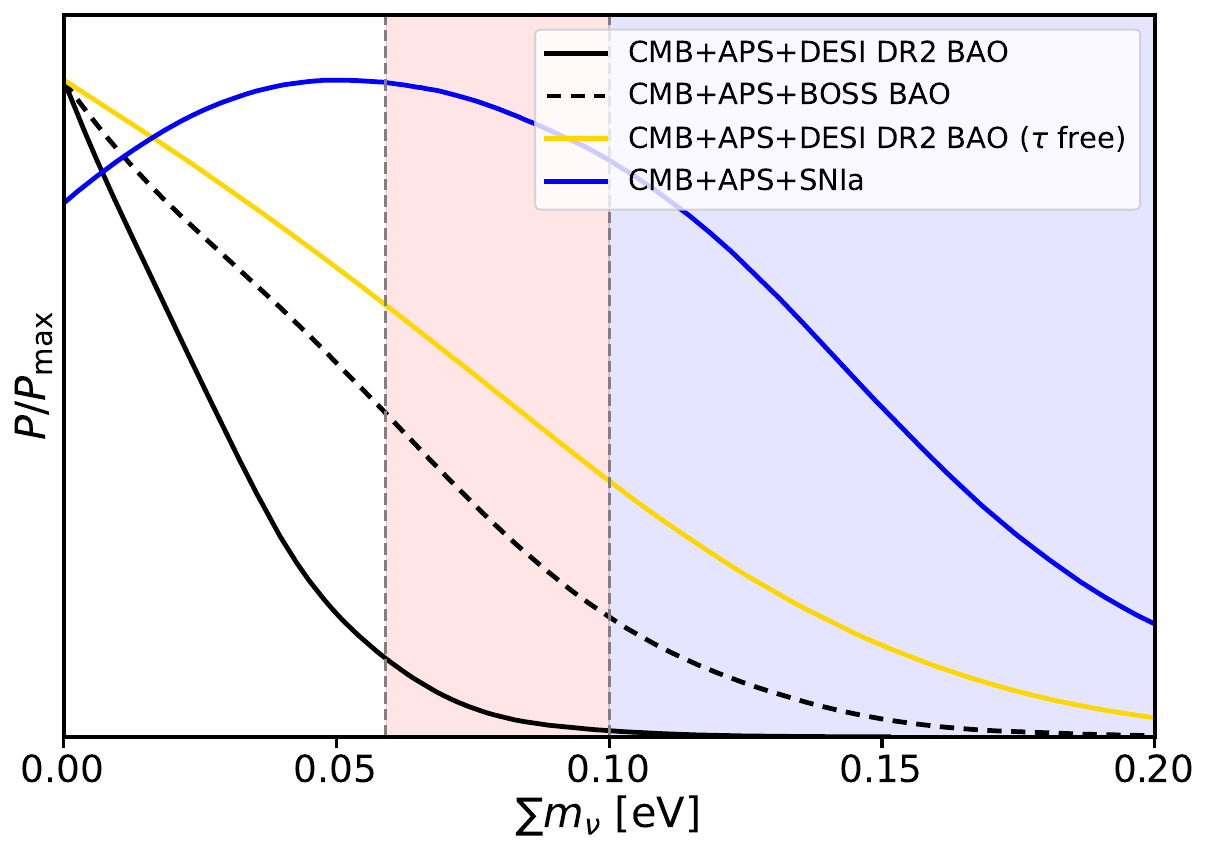}
    \caption{  $95\%$ upper limits on the sum of the neutrino masses, $\sum{m_\nu}$, within the $\Lambda$CDM model. Reference~\citep{desicollaboration2025desidr2resultsi} showed that this distribution is broadened when relaxing the assumption of the dark energy being a cosmological constant. The limits also become more relaxed when replacing DESI DR2 BAO with BOSS BAO (dashed black), \texttt{Pantheon+} SNIa (blue) or when not relying on the optical depth $\tau$  by removing \textit{Planck} low-$\ell$ EE (orange). The vertical dashed lines and shaded regions indicate the minimum allowed $\sum{m_\nu}$ values for the normal and inverted mass-ordering scenarios.}
    \label{fig:mnu} 
\end{figure}

Figure~\ref{fig:mnu} summarizes some constraints on the neutrino mass sum based on different dataset choices. Replacing DESI DR2 BAO with BOSS BAO, one obtains
\begin{equation}
    \Sigma m_\nu < 0.112 \ \mathrm{eV \ (95\% \ \mathrm{C.L.},  \ \textsf{CMB} + APS +\mathrm{BOSS} \ BAO} ) .  
\end{equation}
Excluding  BAO completely and instead using supernova measurements relaxes the bounds further and results in
\begin{equation}
    \Sigma m_\nu < 0.193 \ \mathrm{eV \ (95\%~ \mathrm{C.L.}, \ \textsf{CMB} + \ APS+ \texttt{Pantheon+}} ),
\end{equation}
which is primarily driven by a higher $\Omega_m$ preferred by the \texttt{Pantheon+} sample.

The neutrino mass sum is also degenerate with the reionization optical depth $\tau$ \cite{Craig:2024tky, Green:2024xbb}.  We can test the sensitivity of our constraints to our knowledge of reionization by excluding \textit{Planck} low-$\ell$ EE, i.e., \texttt{SRoll2}, resulting in a relaxed upper bound of
\begin{equation}
    \Sigma m_\nu < 0.150 \ \mathrm{eV \ (95\%~ C.L.},\textsf{CMB} + \ \mathrm{APS} + \ \mathrm{BAO}, \tau~\textrm{free}).
\end{equation}

So far, all neutrino mass constraints are derived within $\Lambda\mathrm{CDM}+\sum m_\nu$. 
However, some models, such as ones that allow the dark energy equation of state to change with time~\citep{desicollaboration2025desidr2resultsi,garciaquintero2025cosmologicalimplicationsdesidr2}, introduce parameters that are degenerate with $\Omega_mh^2$, opening up different ways to relax constraints on $\sum m_\nu$.
We defer more exhaustive studies aimed at discerning the impact of data and model choices on constraints of the neutrino mass sum to future work.

\vspace{0.2cm}
\section{Discussion}
We have presented cosmological constraints from the first joint analysis of CMB lensing from ACT DR6, \textit{Planck} PR4 and \sptshort. 
Building on previous separate analyses of these datasets, we release a joint lensing likelihood and provide tight constraints on the amplitude of density fluctuations on mainly linear scales in the redshift range $z\approx0.9$--$5$.  In the $\Lambda$CDM framework, we constrain $S^{\mathrm{CMBL}}_8$ to $1.6\%$ and provide a $1.1\%$ determination of $\sigma_8$ when combined with BAO.

The amplitudes of structure growth inferred from the three different experiments are fully consistent with each other, with each individual experiment constraining $S^{\mathrm{CMBL}}_8$ at around the $2\%$ level. Our results are in excellent agreement with the model predictions from the $\Lambda$CDM fits to the \textsf{CMB} of P-ACT and reinforce the conclusions of \cite{ACT:2023ubw,ACT:2023kun,farren2025atacamacosmologytelescopemultiprobe,SPT-3G:2024atg} that structure growth follows $\Lambda$CDM expectations over a broad range of scales and redshifts.



In addition, using BAO, we provide a $0.8\%$ constraint on the Hubble constant. We further measure $H_0$ independently of the sound horizon scale and BAO with $4\%$ precision using uncalibrated supernovae. Both methods are in agreement with each other and in agreement with the constraints derived from the primary CMB.  

Finally, with the assumption that the cosmological model is the $\Lambda\mathrm{CDM}+\sum m_\nu$ model, the combination of \textsf{CMB}, BAO and CMB lensing results in an upper bound on the  neutrino mass sum $\Sigma m_\nu < 0.062 \ \mathrm{eV \ (95\% \ C.L.)}$, compared with the physical prior $\Sigma m_\nu  \geq 0.059 \ \mathrm{eV}$ from neutrino oscillation experiments \citep{2019JHEP...01..106E,PhysRevD.110.030001}. Although our results are similar to previously reported constraints, understanding the origin of this tight neutrino mass constraint, including possible inconsistencies between different data types or specific assumptions about the cosmological model, remains important.

Combining multiple surveys probing the same observable has significant potential to improve the constraints of the cosmological parameters. 
We reached the highest signal-to-noise CMB lensing measurement by combining the individual measurements from ACT DR6, \textit{Planck} PR4 and \sptshort.
This methodology not only highlights the current advantages of such synergistic approaches but also paves the way for even greater improvements as new CMB lensing data from ACT DR6+, SPT-3G~\citep{prabhu2024testingmathbflambdacdmcosmologicalmodel}, the Simons Observatory (SO)~\citep{2019JCAP...02..056A}, CMB-S4~\citep{CMB-S4} and CMB-HD~\citep{CMB-HD} become available.

\section{Data Availability}
The MCMC chains used in this analysis are publicly available~\cite{APSLensingChains}. The likelihood code is available~\cite{SPTACTLikelihood}.


WLKW acknowledges support from an Early Career Research Award of the Department of Energy and a Laboratory Directed Research and Development program as part of the Panofsky Fellowship program at the SLAC National Accelerator Laboratory. 
The SLAC authors acknowledge support by the Department of Energy, under contract DE-AC02-76SF00515. 
IAC acknowledges support from Fundaci\'on Mauricio y Carlota Botton and the Cambridge International Trust. 
BDS acknowledges support from the European Research Council (ERC) under the European Union’s Horizon 2020 research and innovation programme (Grant agreement No. 851274). BDS further acknowledges support from an STFC Ernest Rutherford Fellowship.
NS acknowledges support from DOE award number DE-SC0025309.
AC acknowledges support from the STFC (grant numbers ST/W000977/1 and ST/X006387/1).
JD acknowledges support from NSF grant AST-2108126, from a Royal Society Wolfson Visiting Fellowship and from the Kavli Institute for Cosmology Cambridge and the Institute of Astronomy, Cambridge. 
OD acknowledges support from a SNSF Eccellenza Professorial Fellowship (No. 186879).
CS acknowledges support from the Agencia Nacional de Investigaci\'on y Desarrollo (ANID) through Basal project FB210003.
KM acknowledges support from the National Research Foundation of South Africa.
JC acknowledges support from a SNSF Eccellenza Professorial Fellowship (No. 186879).
R.D. thanks ANID for grants BASAL CATA FB210003, FONDEF ID21I10236 and QUIMAL240004

Support for ACT was through the U.S.~National Science Foundation through awards AST-0408698, AST-0965625, and AST-1440226 for the ACT project, as well as awards PHY-0355328, PHY-0855887 and PHY-1214379. Funding was also provided by Princeton University, the University of Pennsylvania, and a Canada Foundation for Innovation (CFI) award to UBC. The development of multichroic detectors and lenses was supported by NASA grants NNX13AE56G and NNX14AB58G. Detector research at NIST was supported by the NIST Innovations in Measurement Science program. 
ACT operated in the Parque Astron\'omico Atacama in northern Chile under the auspices of the Agencia Nacional de Investigaci\'on y Desarrollo (ANID). We thank the Republic of Chile for hosting ACT in the northern Atacama, and the local indigenous Licanantay communities whom we follow in observing and learning from the night sky.

The South Pole Telescope program is supported by the National Science Foundation (NSF) through awards OPP-1852617 and OPP-2332483. Partial support is also provided by the Kavli Institute of Cosmological Physics at the University of Chicago.

Argonne National Laboratory’s work was supported by the U.S. Department of Energy, Office of High Energy Physics, under contract DE-AC02-06CH11357.  Work at the Fermi National Accelerator Laboratory (Fermilab), a U.S. Department of Energy, Office of Science, Office of High Energy Physics HEP User Facility, is managed by Fermi Forward Discovery Group, LLC, acting under Contract No. 89243024CSC000002.

Computations were performed on the Niagara supercomputer at the SciNet HPC Consortium. SciNet is funded by Innovation, Science and Economic Development Canada; the Digital Research Alliance of Canada; the Ontario Research Fund: Research Excellence; and the University of Toronto. This research also used resources of the National Energy Research Scientific Computing Center (NERSC), a DOE Office of Science User Facility supported by the Office of Science of the U.S. Department of Energy under Contract No. DE-AC02-05CH11231
\bibliography{apssamp}

\section{Likelihood Implementation Details}\label{app.likelihood_details}In this section, we present the details of how the joint lensing likelihood is implemented.
\vspace{-20pt}
\subsection{Covariance Matrix Estimation}
\vspace{-15pt}
The main diagonal of the covariance matrix $\mathbb{C}_{b b'}$ corresponds to the CMB lensing auto-covariance estimated with simulations provided by the respective experiments for ACT and \textit{Planck} lensing data. The SPT-3G auto-covariance is derived following \citet{SPT-3G:2024atg}. The cross-covariance between ACT and \textit{Planck} $\mathbb{C}_{b b'}^{\kappa_A\kappa_A,\kappa_P \kappa_P}$ is estimated using a set of 480 \texttt{FFP10} CMB simulations on the ACT footprint \citep{ACT:2023dou}.

Detailed calculations of the approximate off-diagonal correlation coefficients between ACT and SPT $\mathbb{C}_{b b'}^{\kappa_A\kappa_A,\kappa_S \kappa_S}$ ($\lesssim15\%$) and \textit{Planck} and SPT $\mathbb{C}_{b b'}^{\kappa_P\kappa_P,\kappa_S \kappa_S}$ ($\lesssim10\%$) are provided in the following section. The effect of including these off-diagonal terms is small since the SPT overlap with ACT and with \textit{Planck} is a small fraction of both the ACT and \textit{Planck} sky areas. Moreover, the CMB lensing of \sptshort\ is entirely derived from CMB polarization signal, while the CMB temperature signal dominates the information in ACT and {\it Planck} CMB lensing reconstructions. Furthermore, the lensing bandpowers of \sptshort\ have a higher signal-to-noise ratio on smaller lensing scales compared to ACT and \textit{Planck}.
\vspace{-20pt}
\subsection{MCMC Implementation}
\vspace{-15pt}
We consider MCMC chains to be converged if the Gelman--Rubin statistic \citep{10.1214/ss/1177011136} satisfies $R-1 \leq 0.01$. 

\vspace{-20pt}
\subsection{Theory Spectrum Calculation}
\vspace{-15pt}
We model the CMB lensing spectrum using the non-linear matter power spectrum from \texttt{HALOFIT}~\citep{Mead:2016zqy}. The three lensing likelihoods consistently handle the theory bandpowers: they are the bandpower window functions from each experiment multiplied to the output from \texttt{CAMB}/\texttt{class\_sz}. Specifically, the ACT and \textit{Planck} lensing-only measurements (available at \href{https://github.com/ACTCollaboration/act_dr6_lenslike}{this Github repository}) do not have likelihood corrections applied to theory predictions for runs that do not include primary CMB data. The \sptshort\ lensing bandpowers are primary CMB and systematics marginalized and do not require standard QE lensing-only likelihood corrections. We have verified that the posterior distributions are robust to the \texttt{HALOFIT} version used and to variations in the prescriptions of baryonic effects via \texttt{HMCODE}\citep{Mead:2020vgs}.

\vspace{-20pt}
\subsection{Priors Used}\label{app.prior}
\vspace{-15pt}
Table~\ref{table:priors} shows the priors used in the lensing only cosmological analysis.
\begin{table}[h!]
\centering
\begin{tabular}{cc}
\hline\hline
Parameter       & Prior      \\ \hline
$\ln (10^{10}A_s)$ & $[2,4]$           \\ 
$H_0 [\hun]$           & $[40,100] $        \\ 
$n_s$           & $\mathcal{N}(0.96,0.02)$     \\ 
$\Omega_bh^2$   & $\mathcal{N}(0.0223,0.0005)$ \\ 
$\Omega_ch^2$   & $[0.005,0.99]$    \\ 
$\tau$          & $0.055$  (fixed)         \\ \hline
\end{tabular}
\caption{Priors used in the lensing-only cosmological analysis of this work. Uniform priors are shown in square brackets and Gaussian priors with mean $\mu$ and standard deviation $\sigma$ are denoted $\mathcal{N}(\mu,\sigma)$. The priors adopted here
 are identical to those used in the lensing power spectrum analyses of \textit{Planck}~\citep{Planck:2018} and ACT DR6~\citep{ACT:2023kun}. The prior on the baryon density $\Omega_bh^2$ used here is inferred from deuterium abundance measurements and their Big Bang Nucleosynthesis (BBN) predictions \citep{10.1038/s41586-020-2878-4}.
 }
\label{table:priors}
\end{table}

\vspace{20pt}

\section{Analytic estimate of the covariance matrix}\label{app.covmat}
We estimate the joint covariance between SPT-3G and \textit{Planck} CMB lensing.  Correlations arise not only from the shared underlying gravitational potential being measured, but also from the reconstruction noise $n_0$ since part of this is sourced by the same CMB fluctuations that mimic lensing. Given that the full SPT-3G footprint is enclosed within the \textit{Planck} footprint, we decompose the \textit{Planck} convergence field as follows: 
\begin{equation}
    \kappa^{P}(\hat{n}) 
= 
\begin{cases}
 \kappa_{\mathrm{in}}(\hat{n}) + n^P_{{\mathrm{in},0}}(\hat{n}), 
       & \text{inside SPT patch},\\[6pt]
   \kappa_{\mathrm{out}}(\hat{n}) + n^P_{{\mathrm{out},0}}(\hat{n}), 
       & \text{outside SPT patch}.
\end{cases}
\end{equation}
Here $\kappa_{\mathrm{in}}(\hat{n})$ is the lensing field on the SPT overlap region, $\kappa_{\mathrm{out}}(\hat{n})$ is on the rest of \textit{Planck} sky and $n^P$ denotes the \textit{Planck} reconstruction noise.

For SPT-3G we have 
\begin{equation}
    \kappa^{S}(\hat{n}) 
= 
\begin{cases}
   \kappa_{\mathrm{in}}(\hat{n}) + n^S_{{\mathrm{in},0}}(\hat{n}), 
       & \text{inside SPT patch},\\[6pt]
   0  
       & \text{outside SPT patch}.
\end{cases}
\end{equation}

From the definitions above, it is clear that only the overlapping region between both surveys contributes to their correlation and thus the covariance matrix. For simplicity, we drop the subscript "in" going forward, noting explicitly that $\kappa(\hat{n})$ and $n_0$ refers exclusively to the overlap region between SPT-3G and \textit{Planck}.

For the analytic covariance between the two measured lensing spectra $\hat{C}^{\kappa_P\kappa_P}_L$ and $\hat{C}^{\kappa_S\kappa_S}_L$, we have from Wick's theorem
\begin{equation}
    \mathrm{Cov}(\hat{C}^{\kappa_P\kappa_P}_L,\hat{C}^{\kappa_S\kappa_S}_L)=\frac{2}{(2L+1)f_\mathrm{\mathrm{sky,overlap}}}(\hat{C}^{\kappa_P\kappa_S}_L)^2.
\end{equation}

The correlation coefficient within the overlap for the two surveys is then given by
\begin{equation}
    \rho^{\mathrm{overlap}}_L(P,S)=\frac{\mathrm{Cov}(\hat{C}^{\kappa_P\kappa_P}_L,\hat{C}^{\kappa_S\kappa_S}_L)}{\sqrt{\mathrm{Var(\hat{C}^{\kappa_P\kappa_P}_L)}}\sqrt{\mathrm{Var(\hat{C}^{\kappa_S\kappa_S}_L)}}}
\end{equation}

The denominator can be obtained from the raw non-bias subtracted reconstructions of the different instruments in the overlap region. To calculate the full correlation coefficient accounting for the decorrelation due to the difference in area, one simply needs to rescale $\rho^{\mathrm{overlap}}_L(P,S)$ accounting for the different $f_\mathrm{sky}$ of the different footprints.
\begin{equation}
    \rho^{}_L(P,S)=\frac{f^\mathrm{overlap}_\mathrm{sky}}{\sqrt{f^\mathrm{S}_\mathrm{sky}f^\mathrm{P}_\mathrm{sky}}}\rho^{\mathrm{overlap}}_L(P,S)
\end{equation}

\begin{figure}[h!]
    \centering    \includegraphics[width=\columnwidth]{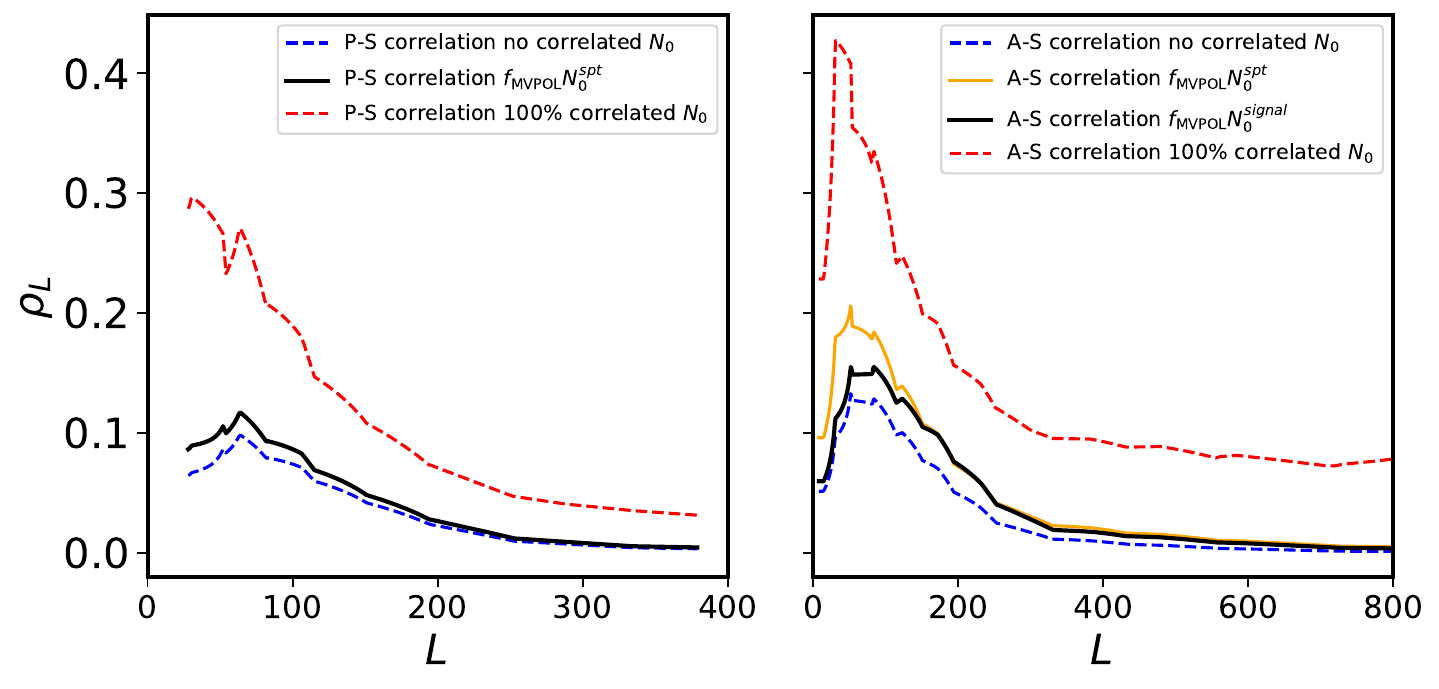}
    \caption{\textbf{Left}: Correlation coefficient estimates between \textit{Planck} and SPT-3G. In solid black, we show the analytic correlation coefficient used to generate the off-diagonal covariance matrix between \textit{Planck} and SPT. We also show in the dashed blue line an underestimate of this correlation coefficient when the correlated part of the reconstruction noise between the two experiments is not included. In red we show an unrealistic pessimistic estimate of the correlation coefficient where we assume the reconstruction noise of both surveys is fully correlated.
    \textbf{Right}:  Analogous estimates for ACT DR6 and SPT-3G.  Note that in this case, the analytic correlation coefficient used (black) enables us to use the ACT MVPOL RDN0 (that only has the signal contribution to the reconstruction noise) as an estimate of the true correlation between the two surveys' reconstruction noise; as expected, this is lower than the orange curve (calculated similarly to the black curve in the left plot) in the signal-dominated regime. }
    \label{fig. corr_coeff}
\end{figure}

For the numerator $\hat{C}^{\kappa_S\kappa_P}_L$,
if we have assumed that the reconstruction noise is uncorrelated between SPT and \textit{Planck}, 
$N_L^{\kappa_P\kappa_S} \;\equiv\; \bigl\langle n^P\,n^S \bigr\rangle_L = 0$, then it is just $(C^{\kappa\kappa}_L)^2$ without any reconstruction noise, corresponding to the lower bound in the estimate of the correlation coefficient shown in  \textcolor{blue}{dashed blue} in the left panel of Fig.~\ref{fig. corr_coeff}.

However, we do expect the signal component of the \textit{Planck} polarization only (MVPOL) reconstruction noise to be correlated with the signal part of the SPT-3G reconstruction noise (which is polarization only) due to common CMB modes. If one naively assumes that all the signal part of the \textit{Planck} reconstruction noise bias ($N_0$) correlates with the SPT-3G $N_0$, then $\hat{C}^{\kappa_P\kappa_S}_L={C}^{\kappa\kappa}_L+N^{SPT}_L$, i.e., the reconstruction noise power spectrum from the SPT-3G measurement. For illustration, this is shown as the \textcolor{red}{dashed red} curve in the left panel of Fig.~\ref{fig. corr_coeff}, which will be an overestimate since the polarization contribution to the \textit{Planck} $\hat{\kappa}$ is subdominant to the temperature channel. A quick estimate of the fraction of the MVPOL lensing reconstruction in \textit{Planck} is given by 
\begin{equation}
    f_\mathrm{MVPOL}=\frac{\sum_{XY\in\{EE,EB\}}\mathcal{R}^{XY}_L}{\sum_{XY}{\mathcal{R}^{XY}_L}}
\end{equation}
i.e. the ratio of the inverse MVPOL normalization to the inverse of the minimum-variance (MV) normalization, where $XY \in [TT, TE, EE, EB, TB]$. Thus the cross-reconstruction noise power spectrum is given by $\langle{n^\mathrm{signal,S}_0+n^\mathrm{noise,S}_0,    f_\mathrm{MVPOL}(n^\mathrm{signal,P}_0}+n^\mathrm{noise,P}_0)\rangle=f_\mathrm{MVPOL}N^\mathrm{signal,SPT}_0\leq{f_\mathrm{MVPOL}N^\mathrm{SPT}_0}$. We approximate the numerator term as $\hat{C}^{\kappa_P\kappa_S}_L={C}^\mathrm{\kappa\kappa}_L+f_\mathrm{MVPOL}{N}^\mathrm{SPT}_L$ noting that this will be an upper bound since the instrument noise part of the SPT $N_0$ will not appear in reality. This upper bound is shown in \textbf{black} in the left panel of Fig. \ref{fig. corr_coeff}.

We repeat the same calculation to estimate the correlation between ACT and SPT-3G. Noting that in this case, for the correlation due to reconstruction noise $\langle{n^\mathrm{signal,S}_0+n^\mathrm{noise,S}_0,    f^\mathrm{A}_\mathrm{MVPOL}(n^\mathrm{signal,A}_0})\rangle=f^\mathrm{A}_\mathrm{MVPOL}N^{ \mathrm{signal,ACT}}_0$, one can use the realization dependent $N_0$ of ACT that does not have contribution from instrument noise to estimate a more accurate correlation due to the signal contribution to the reconstructed noise.

\begin{figure}[h!]
    \centering    \includegraphics[width=0.8\columnwidth]{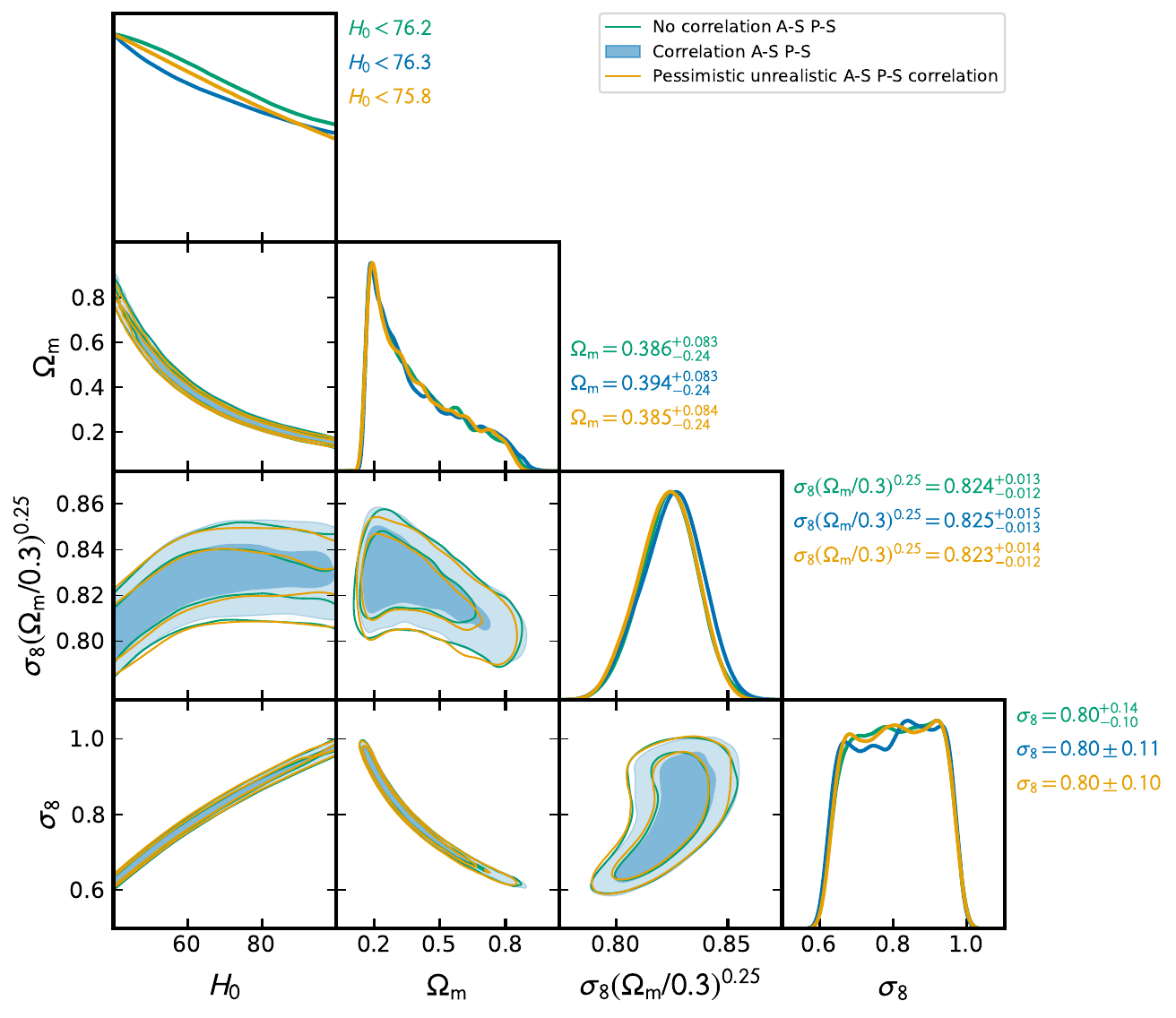}
    \caption{Comparison of parameter constraints obtained under three different treatments of off-diagonal covariance components between SPT-3G M2PM and both \textit{Planck} PR4 and ACT DR6. The green curves show constraints when off-diagonal contributions are omitted entirely, the blue curves correspond to the baseline analytic correlation model and the orange curves depict a pessimistic scenario assuming full correlation for overlapping bins. Even in the pessimistic case, the impact of including off-diagonal terms is minimal compared to neglecting them.}
    \label{fig.comparison}
\end{figure}

We convert the analytical estimates of the correlation coefficients into the off-diagonal elements of the covariance matrix. Although these analytic estimates do not explicitly account for correlations between non-overlapping bins, we only expect significant correlations for bins where the bandpowers of the different surveys overlap in multipole space. For overlapping bins, we assign a correlation coefficient, $\rho$, at the multipole corresponding to the average of the two survey bin centres. For bin pairs whose centers differ by more than 50, we set $\rho=0$, since we do not expect any significant correlation between non-overlapping bins. (The use of realization dependent $N_0$ ensures that off-diagonal correlations are kept at a minimum \citep{PhysRevD.83.043005,Schmittfull_2013}.) In Figure~\ref{fig.comparison}, we compare parameter constraints under three scenarios: excluding off-diagonal contributions between SPT and both \textit{Planck} and ACT (green), using the baseline analytic correlations (blue) and employing the unrealistic pessimistic correlations described above (orange). Notably, even in the pessimistic unrealistic scenario, including the off-diagonal components has only a minimal effect compared to omitting them entirely.

We further verify, using a simulation approach, that our analytic estimates are reliable. We generate CMB skies consistent with those used by the SPT-3G survey and obtain lensing reconstructions from them. The conclusions from the simulated covariance matrix are the same as for the analytic case: correlations between Planck and SPT are negligible, while those between neighboring bins in the ACT--SPT block are at the 10--20\% level. Given the noisier estimates obtained with the simulation-based covariance matrix, as seen in the off-diagonal blocks where no physical correlation is expected, we adopt the analytic covariance as our estimate of the correlations.

\begin{figure}[h!]
    \centering    \includegraphics[width=\columnwidth]{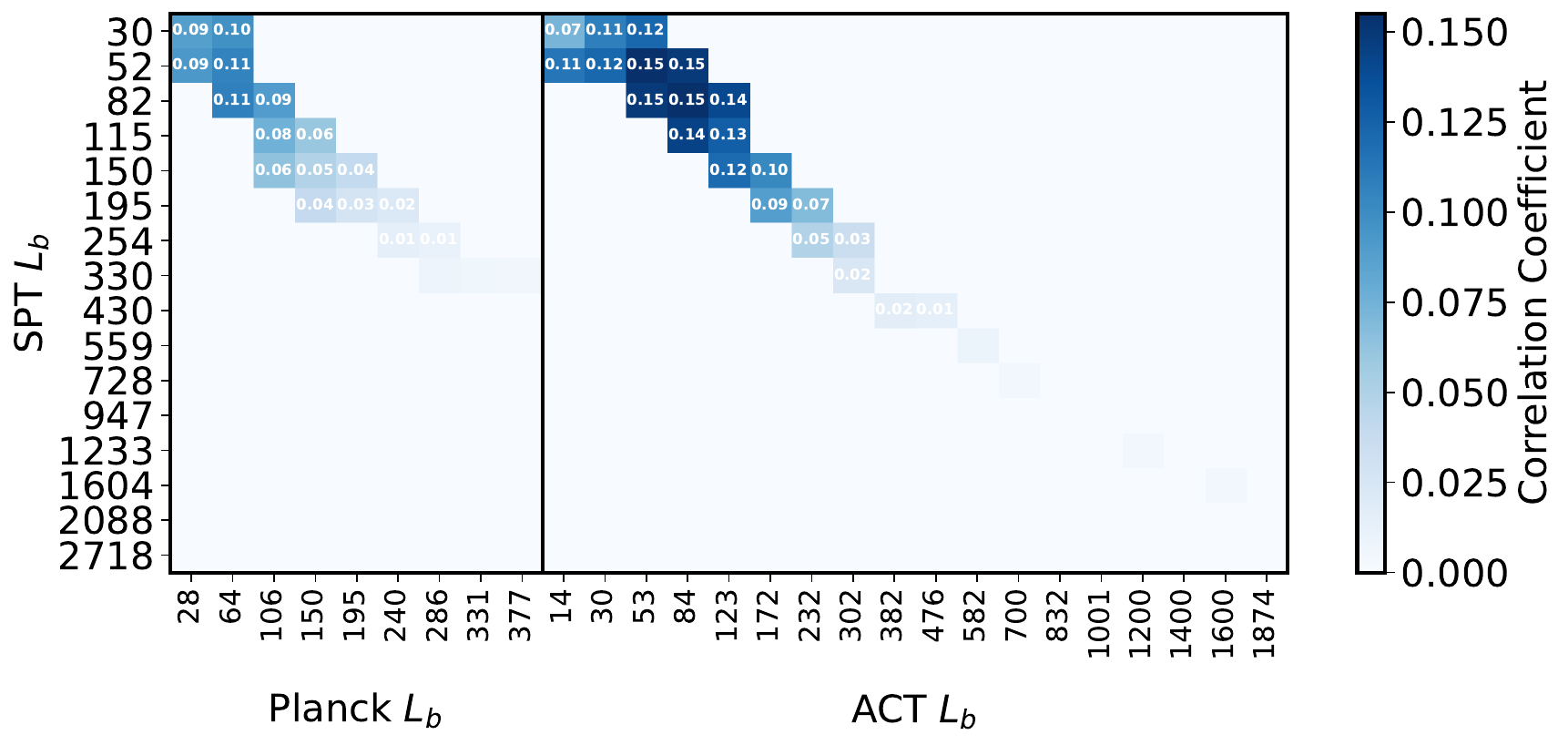}
    \caption{Binned correlation coefficient for ACT-SPT-3G and \textit{Planck}-SPT-3G used in the covariance matrix of the likelihood.}
    \label{fig. corr_coeff_matrix}
\end{figure}

\vspace{20pt}

\section{Results using alternative BAO datasets}\label{app.variation}
We explore three variants of BAO datasets in this analysis. They are Pre-DESI BAO, DESI DR1 BAO  and Hybrid BAO, defined as follows. 
Pre-DESI BAO and Hybrid BAO are included for comparisons with the results produced in the ACT DR6 lensing~\cite{ACT:2023dou,ACT:2023kun,ACT:2023ubw} and the \sptshort~\citep{SPT-3G:2024atg} papers, respectively. The parameter constraints from CMB lensing with these BAO datasets are samarized in Table~\ref{tab:params2}.
\begin{itemize}
    \item Pre-DESI BAO consists of BAO  measurements from 6dFGS \cite{2011MNRAS.416.3017B}, SDSS DR7 MGS \cite{1409.3242}, BOSS DR12 LRGs \cite{1607.03155} and eBOSS DR16 LRGs \cite{2007.08991}. 
    \item DESI DR1 BAO denotes the DESI-Y1 release \cite{desicollaboration2024desi}, covering redshifts $0.1\leq{z}\leq4.2$ with samples including BGS, LRG, ELG, QSO and the Lyman-$\alpha$ forest.
    \item Hybrid BAO is based on the DESI BAO set but replaces the DESI BGS and lowest redshift DESI LRG with SDSS MGS $(z_\mathrm{eff}\sim0.15)$ and two BOSS DR12 LRG points at $(z_\mathrm{eff}\sim0.38,0.51)$. The DESI Ly$\alpha$ point is also replaced with the joint eBOSS+DESI DR1 Ly$\alpha$ BAO. The 6dFGS BAO measurement is also included here. 
\end{itemize}

In combination with APS lensing, they result in the following $\sigma_8$ constraints
\begin{eqnarray}
    \sigma_8 &=& 0.827 \pm 0.010\\ \nonumber &&({\rm ACT}+\textit{Planck}+{\rm SPT}+ {\rm hybrid\, BAO}),\\
    \sigma_8 &=& 0.821 \pm 0.010\\ \nonumber &&({\rm ACT}+\textit{Planck}+{\rm SPT}+ {\rm pre\text{-}DESI\, BAO})\text{\ and}\\
    \sigma_8 &=& 0.831 \pm 0.010\\ \nonumber &&({\rm ACT}+\textit{Planck}+{\rm SPT}+ {\rm DESI \,DR1\, BAO}).
\end{eqnarray}

From the combination of the joint CMB lensing,  galaxy BAO and the $\Omega_bh^2$ prior in Table~\ref{table:priors}, we obtain a $1.1$--$1.4\%$ constraint on $H_0$:
\begin{eqnarray}
    H_0 &=& 68.46\pm 0.73 \hun\\ \nonumber &&({\rm ACT}+\textit{Planck}+{\rm SPT}+ {\rm hybrid\, BAO}),\\
    H_0 &=& 68.42\pm 0.98 \hun\\ \nonumber &&({\rm ACT}+\textit{Planck}+{\rm SPT}+ {\rm pre\text{-}DESI\, BAO})\text{\ and}\\
    H_0 &=& 69.15\pm 0.74 \hun\\ \nonumber &&({\rm ACT}+\textit{Planck}+{\rm SPT}+ {\rm DESI \,DR1 \, BAO}).
\end{eqnarray}

\begin{table}[ht]
\begin{tabular}{l  c  c  c } \hline\hline
Experiment & $S^{\mathrm{lens}}_8$ & $\sigma_8$ & $\Omega_m$ \\
\hline
A+ pdBAO & $0.829 \pm 0.020$ & $0.819 \pm 0.015$ & $0.315 \pm 0.016$ \\
S+ pdBAO  & $0.836 \pm 0.012$ & $0.821 \pm 0.013$ & $0.323 \pm 0.018$ \\
{P} + pdBAO & $0.822\pm0.021$  & $0.814\pm0.016$ &  $0.313^{+0.014}_{-0.017}$\\
AS+pdBAO & $0.836 \pm 0.011$ & $0.823 \pm 0.011$ & $0.319 \pm 0.014$ \\
APS+ pdBAO & $0.828 \pm 0.010$ & $0.831 \pm 0.010$ & $0.317 \pm 0.012$ \\
APS+ DESI DR1 BAO & $0.832 \pm 0.010$ & $0.821 \pm 0.010$ & $0.296 \pm 0.010$ \\
APS+ hBAO & $0.828^{+0.011}_{-0.010}$ & $0.827 \pm 0.010$ & $0.302^{+0.009}_{-0.010}$ \\
APS+ DESI DR2 BAO & $0.829^{+0.009}_{-0.009}$ & $0.829 \pm 0.009$ & $0.300\pm0.007$\\
\hline
\end{tabular}
\caption{Cosmological parameter measurements from the various lensing experiment combinations. We use A, {P} and S as shorthands for CMB lensing with ACT DR6, \textit{Planck} PR4 and \sptshort\ respectively. pdBAO refers to pre-DESI BAO and hBAO stands for hybrid BAO.}
\label{tab:params2}
\end{table}

\vspace{20pt}

\section{$\sigma_8$ Constraints from galaxy and CMB lensing}\label{sec:sig8}
\begin{figure}
    \centering
    \includegraphics[width=0.8\linewidth]{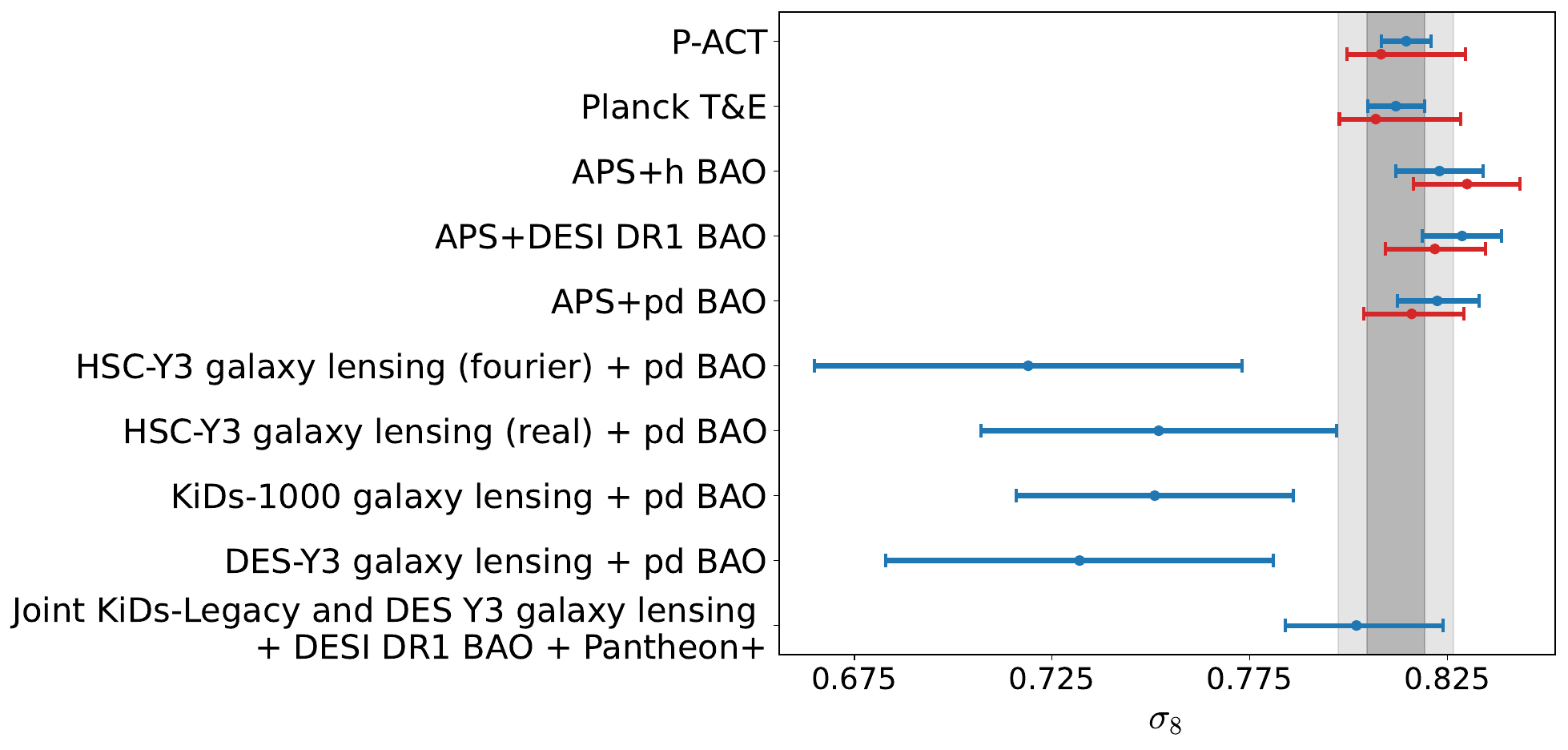}
    \caption{
    Constraints on $\sigma_8$ inferred from galaxy lensing surveys of DES-Y3 \cite{DES:2021bvc, DES:2021vln, DES:2021wwk}, KiDS-1000 \citep{KiDS:2020suj, Heymans:2020gsg} and HSC-Y3 \citep{Li:2023tui, Dalal:2023olq} (plus BAO), as well as KiDs-Legacy \citep{Wright:2025xka, Stolzner:2025htz} (plus BAO and supernovae),  CMB lensing (plus BAO) and CMB primary spectra from \textit{Planck} PR3 and \textsf{P-ACT}. 
    The blue errorbars show the constraints under the $\Lambda\mathrm{CDM}$ model, while the red errorbars are inferred for a $\Lambda\mathrm{CDM}+\sum m_\nu$ model. The constraints derived from \textit{Planck} 2018 \texttt{TTTEEE lowl lowE} are labeled as Planck T\&E. The vertical gray band also shows the 1 and 2$\sigma$ regions inferred from Planck T\&E assuming $\Lambda\mathrm{CDM}$.
    }
    \label{fig:sigma8}
\end{figure}

In Fig.~\ref{fig:sigma8}, we compare the $\sigma_8$ constraints from galaxy and CMB lensing in combination with BAO measurements. Since CMB lensing measurements extend a large range of angular scales, while the galaxy lensing is mostly sensitive to the structure growth at small scales, CMB lensing is more constraining than galaxy lensing on $\sigma_8$ \cite{ACT:2023kun}.

\end{document}